\documentstyle[psfig,graphics]{mn}
\begin{document}

\title{Time Scales in Long GRBs}

\author[Ehud Nakar \& Tsvi Piran]
{Ehud Nakar \& Tsvi Piran\\
  Racah Institute, Hebrew University, Jerusalem 91904, Israel}

\maketitle
\begin{abstract}
We analyze a sample of bright long bursts and find that the pulses duration
have a lognormal distribution while the intervals between pulses have an excess
of long intervals (relative to lognormal distribution). This excess can be explained
by the existence of \textit{quiescent times}, long periods with no signal above
the background. The lognormal distribution of the intervals (excluding the \textit{quiescent
times}) is similar to the distribution of the pulses width. This result suggests
that the \textit{quiescent times} are made by a different mechanism than the
rest of the intervals. It also suggests that the intervals (excluding the \textit{quiescent
times}) and the pulse width are connected to the same parameters of the source.
We find that there is a correlation between a pulse width and the duration of
the interval \textit{preceding} it. There is a weaker, but still a significant,
correlation between a pulse width and the interval \textit{following} it. The
significance of the correlation drops substantially when the intervals considered
are not adjacent to the pulse.
\end{abstract}


\section{Introduction}

The light curve of a long duration Gamma-Ray Burst (GRB) (Kouveliotou 1993) is usually complex, it is composed
from several dozens of short (about 1sec) distinguishable pulses. The variability
of these bursts has been already used (Sari \& Piran, 1997; Fenimore, Ramirez
\& Sumner 1997) to suggest that internal shocks , rather than external shocks
produce most GRBs (simple models of the latter cannot produce efficiently variable
bursts). The recent discovery of correlation relating properties of
the temporal structure with the burst luminosity offer possibility of deriving
independent estimates of the redshift of GRBs (Stern, Poutanen \& Svensson 1997,
1999; Norris , Marani \& Bonnell 2000; Fenimore \& Ramirez-Ruiz 2000; Reichart
et al 2000).

The temporal properties of long duration GRBs
 , and specifically the pulses and intervals
properties, were investigated in previous works (e.g. McBreen
et.al. 1994, Li \& Fenimore 1996, Norris et. al. 1996, Quilligan
et. al. 1999). We present here a further study of the temporal
properties of GRB light curves. Using a new algorithm (Nakar \&
Piran, 2001b), which is based on Li \& Fenimore (1996), we
analyze the distribution of the time intervals between pulses and
the pulses width in long bright bursts. We find that the pulses
width, \( \delta t \), distribution is consistent with a
lognormal distribution. The distribution of intervals between
pulses, \( \Delta t \), becomes lognormal, with a comparable
width to the \( \delta t \) distribution, provided that we
eliminate intervals that contain \textit{quiescent times}.
Namely, long periods with no evidence of photon counts above the
background, between periods of strong \( \gamma  \)-ray emission
(Ramirez \& Merloni 2001, Nakar \& Piran 2001a). In nature such a
deviation from a lognormal distribution occurs when a different
mechanism govern the high end tail of the distribution. The
similarity between the pulses width and the intervals
distributions suggest that both distributions are influenced by
the same (source) physical parameters. In this case, we expect
also a correlation between the pulse width and its adjacent
intervals. We search for this correlation using 12 bursts
which contain enough pulses for such analysis to be applicable.
In all these bursts there is a significant correlation between
the pulse width and the duration of the interval
\textit{preceding} it. A significant correlation between the pulse
width and the \textit{following} interval was found in 60\% of
the bursts. In all the bursts the significance of the correlation
between the pulse width and the \textit{preceding} interval was
higher or equal to this of the \textit{following} interval.

\section{The Algorithm and the Data}

\label{data}We have applied a modified Li \& Fenimore (1996) peak
finding algorithm (Nakar \& Piran 2001b) to a sample of 68 long
bursts (\( T_{90}>2sec \), where \( T_{90} \) is the time required
to accumulate from 5 to 95 per cent of the total counts). This
algorithm searches for individual peaks in BATSE 64ms concatenate
data. In the analysis we consider the sum of all four energy
channels, i.e. E \( > \) 25Kev. This data type includes the
photon counts, in a 64ms time bins, from a few seconds before the
bursts trigger until a few hundred seconds after the trigger. The
minimal \( \Delta t \) and \( \delta t \) are 0.128sec in this
resolution. Each peak the algorithm find corresponds to one
pulse. We determine the pulse width and the interval between two
maxima, which we define as the interval between pulses (Norris
et. al. 1996, Li \& Fenimore 1996). The algorithm also searches
for \textit{quiescent times}. The definition of minimal duration
of a quiescent time is somewhat arbitrary (whether a single time
bin with a count level of the background is a quiescent time or
not). We demand that the average of ten bins would be at the
background level to be considered as a quiescent time. Hence the
minimal quiescent time in our analysis is about 1sec.

\begin{figure}
\resizebox*{0.9\columnwidth}{0.25\textheight}{\includegraphics{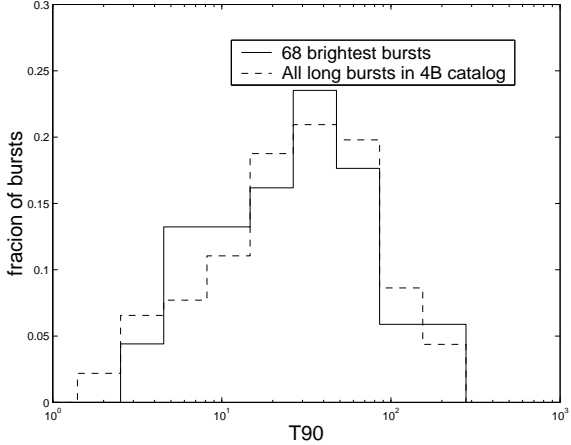}}
\caption{\label{T90 hist fig}The normalized histogram of \( T_{90} \)
of the 68 brightest bursts (solid line), compared to the normalized histogram
of all the long bursts in 4B catalog (dashed line).}
\end{figure}

The sample we consider is the 68 brightest long bursts in the BATSE 4B catalog
(peak flux in 64ms\( > \)10.19\( ph/(sec\cdot cm^{2}) \)). This resulted in
1330 pulses and 1262 intervals. The analysis below is based on the width of
these pulses and the corresponding intervals between them.

One may wonder if the brightest bursts are representative. Figure \ref{T90 hist fig}
depicts the histogram of \( T_{90} \) of our sample compared to the histogram
of all the long bursts in 4B catalog. The histograms are similar. Therefore
our sample, at least as far as duration is concerned, is representative for
the entire 4B catalog.

We also analyze 24 dimmer bursts, which are still bright enough for
the analysis to be carried out. These bursts are chosen randomly from the bursts
with peak flux at 64ms greater than 4\( ph/(sec\cdot cm^{2}) \) but smaller
than 10\( ph/(sec\cdot cm^{2}) \). This sample included 438 intervals. The
results for this sample are consistent with those obtained for the brightest
sample. Again indicating that the brightest sample is representative.

\section{Temporal properties}

\subsection{Pulse Width Distribution}

\begin{figure}
\resizebox*{0.9\columnwidth}{0.25\textheight}{\includegraphics{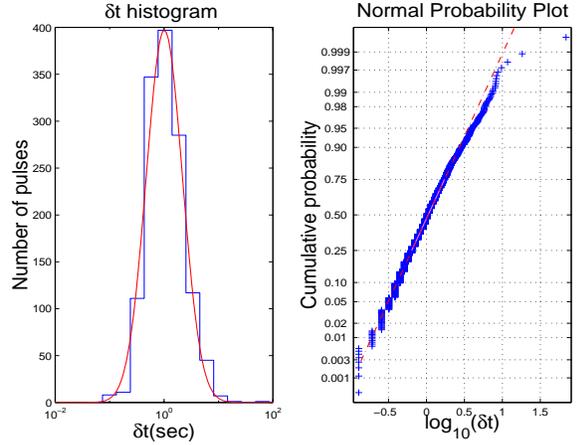}}
\caption{\label{pulse width fig}Left (a):The histogram of the pulses
  width, \protect\( \delta t\protect \),
compared with the best-fit Gaussian. Right (b): The cumulative distribution
of \protect\( log_{10}(\delta t)\protect \) compared with a normal distribution.
Note that the excess of long pulses (top right) is due to a single pulse and
it is not statistically significant.}
\end{figure}

The distribution of the 1330 pulses' widths is in excellent agreement
with a lognormal distribution. The \( \chi ^{2} \) test gives a probability
of \( 0.52 \) that the data was taken from a lognormal distribution, with \( \mu =0.065\pm 0.04 \)
(\( \delta t_{avg}\approx 1sec \)) and \( \sigma =0.77\pm 0.03 \) (corresponding
at \( 1\sigma  \) to 0.5- 2.3sec). Figure \ref{pulse width fig} depicts the
histogram and the cumulative distribution of the pulses' widths in the bright
sample. The analysis of the dimmer sample (see sec \ref{data}) show also an
excellent agreement with a lognormal distribution. The probability that the
dimmer sample's pulses width are taken from a lognormal distribution is 0.9.

\subsection{Intervals Distribution and Quiescent Times}

McBreen (1994) and Li \& Fenimore (1996) suggested that the distribution of
the intervals between pulses, \( \Delta t \), is lognormal. We have found that
the \( \delta t \) distribution is lognormal. These results suggest that we
consider first the null hypothesis that \( \Delta t \) distribution is also
lognormal.

\begin{figure}
\resizebox*{0.9\columnwidth}{0.25\textheight}{\includegraphics{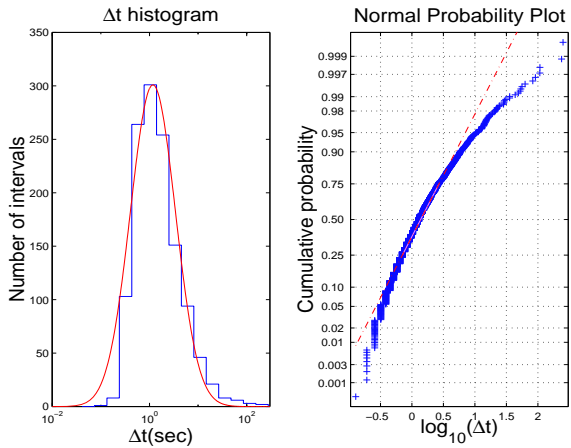}}
\caption{\label{regular interval fig}Left(a):The histogram of the time interval between
neighboring peaks, \protect\( \Delta t\protect \), compared with the best-fit
Gaussian. Right(b): The cumulative distribution of \protect\( log_{10}(\Delta t)\protect \)
compared with a normal distribution.}
\end{figure}

Figure \ref{regular interval fig}a depicts the histogram of the time intervals
between neighboring peaks, \( \Delta t \). Figure \ref{regular interval fig}b
shows the cumulative distribution of \( log_{10}(\Delta t) \), compared to
a best-fit Gaussian. Both figures show a clear excess of many long intervals.
Using the \( \chi ^{2} \) test, we find a probability of \( 1.2\cdot 10^{-10} \)
that the data was taken from a lognormal distribution. The null hypothesis clearly
fails.

\begin{figure}
\resizebox*{0.9\columnwidth}{0.25\textheight}{\includegraphics{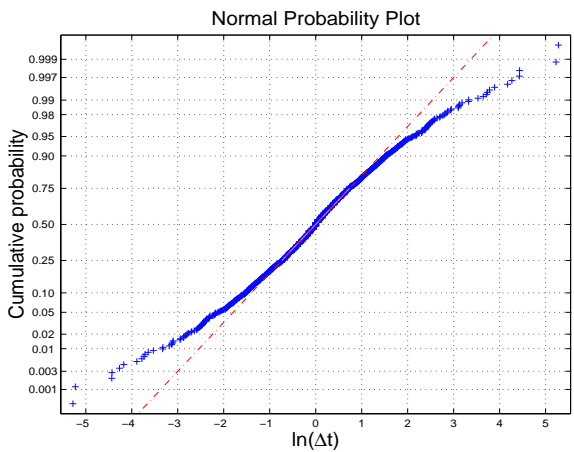}}
\caption{\label{mirror fig}The cumulative distribution of all the
intervals, $\Delta t $, above the median, compared with the best-fit
Gaussian. The intervals above the median form only the right part of the
distribution.
The left half is a duplication of the right one. The entire sample is shifted
by  $ln(median(\Delta t))$ so it is symmetric around zero.}
\end{figure}

Li \& Fenimore (1996) already noticed such a deviation. A hints to this deviation
could also be seen in the results presented by Norris et. al. (1996). Li \&
Fenimore (1996) and McBreen (personal communication 2000) have suggested that
this deviation arises due to the limited resolution of the observations, and
therefore the intervals distribution is consistent with a lognormal one. To
test this hypothesis we examine in Figure \ref{mirror fig} the cumulative probability
of a mirror image of the right half of the \( \Delta t \) histogram. This half
is insensitive to the limited resolution. Again the figure show a deviation
from a lognormal distribution and an excess of long intervals (and short ones
which are of course the mirror of the long intervals). This indicates that \( \Delta t \)
distribution is not a lognormal.

\begin{figure}
\resizebox*{0.9\columnwidth}{0.25\textheight}{\includegraphics{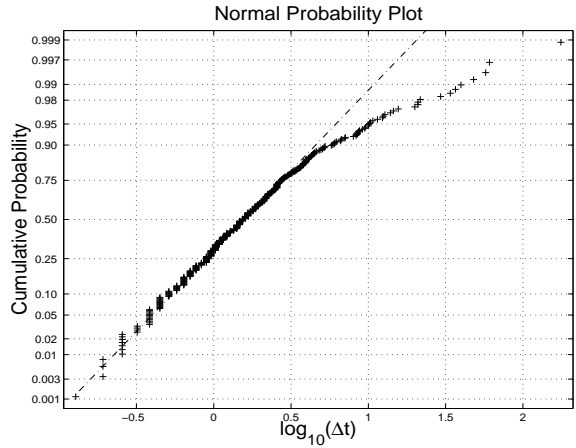}}
\caption{\label{dimDist fig}The cumulative distribution of \protect\( log_{10}(\Delta t)\protect \).
\protect\( \Delta t\protect \) is taken from the sample of 24 dimer bursts.
The cumulative distribution is compared to this of a normal distribution with
the best-fit parameters. }
\end{figure}

The analysis of the dimmer bursts sample (see sec. \ref{data}) gives similar
results. Figure \ref{dimDist fig} shows the cumulative distribution of \( log_{10}(\Delta t) \)
in the dimmer sample compared to the best-fit Gaussian. The excess of long intervals
is clear. The \( \chi ^{2} \) test gives a probability of \( 9\cdot 10^{-6} \)
that this data was taken from a lognormal distribution.

We found that the long intervals between neighboring peaks are often dominated
by \textit{quiescent times}. Such periods are seen in 35 of the bursts in our
sample (all together there are 50 \textit{quiescent times}). Most of the bursts
contain one or two \textit{quiescent times}, some contain three. The \textit{quiescent
times} last typically several tens of seconds and they range between a second
(our arbitrary lower limit) to hundreds of seconds. In some bursts the \textit{quiescent
times} are a significant fraction of the total duration.

\begin{figure}
\resizebox*{0.9\columnwidth}{0.25\textheight}{\includegraphics{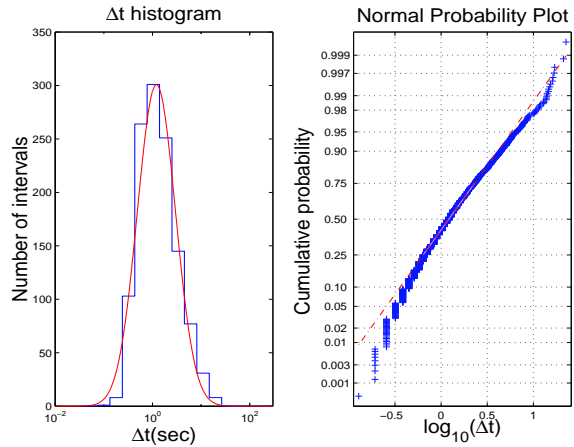}}
\caption{\label{no quiet fig}The histogram and cumulative
distribution of time intervals between neighboring peaks,
\protect\( \Delta t\protect \), excluding intervals that contain
quiescent times, compared with the best-fit Gaussian.}
\end{figure}

Since the \textit{quiescent times} contribute to the duration of the longest
intervals we examined the possibility that they cause the excess of long intervals
(over a lognormal distribution). We examine the interval distribution excluding
now \textit{all} intervals that contained a quiescent time. Figure \ref{no quiet fig}
shows the histogram and the cumulative distribution of \( log_{10}(\Delta t) \)
excluding the intervals that contained a quiescent time, compared to the best-fit
Gaussian with \( \mu =0.257\pm 0.051 \) (\( \Delta t_{avg}\approx 1.3sec \))
and \( \sigma =0.90\pm 0.04 \) (corresponding at \( 1\sigma  \) to 0.53-3.1sec).
The fit is good. The \( \chi ^{2} \) test gives a probability of \( 0.27 \)
that this data was taken from a lognormal distribution. Moreover, the modified
\( \Delta t \) distribution has a comparable width to the \( \delta t \) distribution
(which is unaffected by \textit{quiescent times}).

\subsection{The Correlation Between Pulses and Neighboring Intervals}

The similarity between the two distributions motivated us to explore the correlation
between pulses and their neighboring intervals. We calculated Pearson's linear
correlation coefficient, \( r \), and Spearman's ranked order coefficient,
\( r_{s} \), between a pulse duration and the intervals just preceding it,
and just following it (excluding intervals that contain a quiescent time). In
order to avoid any influence of the algorithm on the correlation we considered
here only well separated pulses, namely, pulses whose half maximum (used to
determine the width) is above the minimum between the pulse and its neighbors.
In this case there is no worry about overlap between the pulses that may introduce
a spurious correlation.

We considered only bursts with more than 12 separated pulses, all together 12
bursts of our sample with an average of 23 pulses per burst\footnote{%
The comparison was done burst by burst in order to eliminate redshift or intrinsic
effects that would have produced spurious correlations if we had considered
the whole data as one set
}. Our null hypothesis is of no correlation between the pulses width and the
intervals. The probabilities of rejection of the null hypothesis found to be
similar for both methods, \( r \) and \( r_{s} \). All the considered bursts
showed a positive correlation between a pulse width and the \textit{preceding}
interval. The rejection of the null hypothesis is above 90\% in half of the
bursts and above 70\% in the rest. The correlation between a pulse width and
the \textit{following} interval is less significant. In 5 of the bursts the
null hypothesis is rejected with a confidence of more then 90\%, and In 7 of
the bursts the rejection confidence level is above 70\%. In 4 bursts there is
no evidence for correlation. In all the considered bursts the significance of
the correlation between the pulse width and the preceding interval is higher
or equal to the correlation with the following interval.

\begin{figure}
\resizebox*{0.9\columnwidth}{0.25\textheight}{\includegraphics{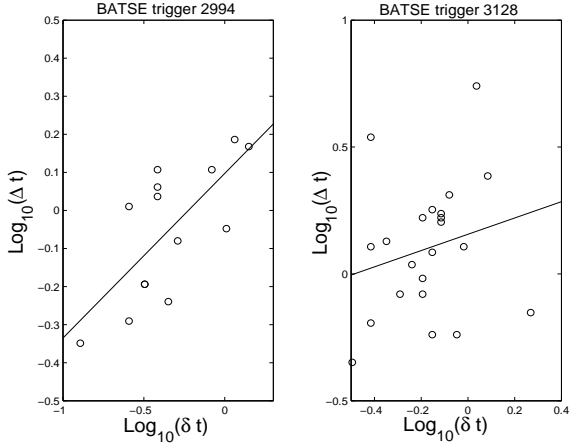}}
\caption{\label{preceed interval correlation}The pulses width and the \textit{preceding}
interval duration in BATSE triggers 2994 and 3128. Trigger 2994 has the most
significant correlation between the two parameters in the analyzed bursts. The
confidence level of this correlation is 99.5\%. Trigger 3128 has the least significant
correlation between the two parameters in the analyzed bursts. The confidence
level of this correlation is 70\%.}
\end{figure}

\begin{figure}
\resizebox*{0.9\columnwidth}{0.25\textheight}{\includegraphics{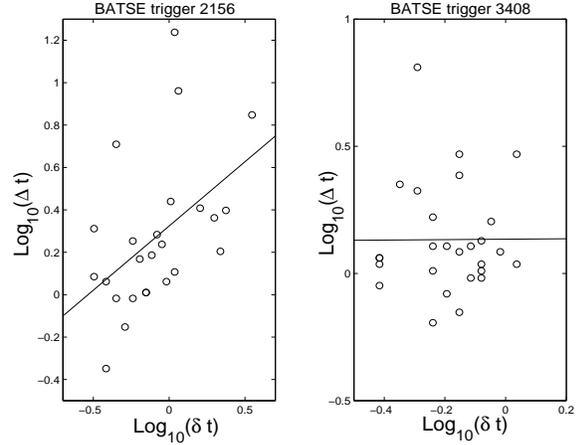}}
\caption{\label{following interval corr}The pulses width and the \textit{following}
interval duration in BATSE triggers 2156 and 3408. Trigger 2156 has the most
significant correlation between the two parameters in the analyzed sample. The
confidence level of this correlation is 98.3\%. There is no evidence for correlation
between the two parameters in trigger 3408.}
\end{figure}

Figure \ref{preceed interval correlation} shows the pulses width and the \textit{preceding}
interval duration of the most correlated burst and the least correlated one.
In both bursts a correlation is clearly seen. The best linear fit parameters
of each burst are different, but the variance of the slope parameter is not
large, compared with the typical slope. The average linear fit is \( log(\Delta t)=(0.49\pm 0.19)\cdot log(\delta t)+(0.25\pm 0.11) \).
Figure \ref{following interval corr} shows the pulses width and the \textit{following}
interval duration of the most correlated burst and the least correlated burst.
While the correlation is clearly seen in trigger 2156, there is no evidence
for correlation in trigger 3408. The average Linear fit for the correlated bursts
(significance level above 70\%) is \( log(\Delta t)=(0.55\pm 0.11)\cdot log(\delta t)+(0.26\pm 0.11) \).

The correlation disappears if we consider non neighboring intervals and pulses.
For example, only half of the considered bursts show a positive correlation
between a pulse duration and the interval after the following pulse. In no burst
the rejection of the null hypothesis of no correlation is above 90\%.

\section{Discussion}

The pulse duration distribution is consistent with a lognormal one. However,
the distribution of the intervals between pulses is inconsistent with a lognormal
distribution. Removal of the intervals that include \textit{quiescent
times} results in a distribution consistent with lognormal. This suggests
that the distribution of interval between pulses is made of the sum of two different
contributions. A lognormal distribution that is similar to the pulse width
distribution, and the \textit{quiescent times} distribution. The last one is
dominant at the high end tail of the intervals distribution and produce the
deviation from a lognormal distribution. We stress (see also Ramirez \& Merloni
2001) that the \textit{quiescent times} that we find here are not the same as
the gaps between the precursors and the main pulses (Koshut et al, 1995) found
in 3\% of the bursts. In that case the gaps are between a softer and weaker
component (the precursor) and a harder and much stronger one (the main burst).
Here we observe gaps between pulses with the same characteristics. Moreover,
in many cases we observe several \textit{quiescent times} within the same burst.

The existence of \textit{quiescent times}, and of two different mechanisms that
control the gaps between pulses pose a new challenge to any model attempting
to explain the temporal structure of GRBs. This suggests that there are three
time scales of different nature within GRBs : (i) The shortest time scale, the
variability scale which determines the pulses duration and the interval between
pulses (both have a similar time-scale). (ii) An intermediate time scale producing
long periods within the bursts with no activity (\textit{quiescent times}).
(iii) The longest time scale which corresponds to the duration of the burst.

Within the popular internal shock model the observed temporal structure reflects
the activity of the {}``inner engine{}'' (e.g. Kobayashi, Piran \& Sari 1997;
Ramirez \& Fenimore, 2000). Within this model \textit{quiescent times} reflect,
most likely, periods in which the {}``inner engine{}'' is not active at all.
Alternatively they could reflect periods in which the {}``inner engine{}''
emits a sequence of shells that do not collide (e.g. shells with a decreasing
Lorentz factors) (Ramirez, Merloni \& Rees 2000). In both cases the \textit{quiescent
times} correspond to periods in which the activity of the {}``inner engine{}''
differs from its usual activity. Within the internal shocks model the shortest
time scale corresponds to the variability of the {}``inner engine{}'', the
\textit{quiescent periods} time scale corresponds to periods of different activity
of the {}``inner engine{}'' and the duration of the burst corresponds to the
total duration of activity of the {}``inner engine{}''.

We also find that the pulse duration is correlated to the preceding interval
duration in all the analyzed bursts. A weaker but still significant correlation
is found between the pulse duration and the following interval duration. This
correlation was found only in 60\% of the analyzed bursts.

While, different mechanisms can produce the observed lognormal distributions,
the similarity between the pulse and interval distributions and the correlation
between the pulses and neighboring intervals require that both processes would
be determined by the same mechanism (or at least by two mechanisms governed
by the same (source) physical parameters). This provides a new constraint
on GRB models that should be considered in any model that attempts to reconstruct
GRBs light curves.

For example, our results should be taken into account when choosing parameters
in simulations of internal shocks light curves. Specifically, the parameters
that determine the temporal characteristics of the relativistic wind should
be chosen such that the resulting pulses and intervals would have lognormal
distributions, with the appropriate parameters as well as \textit{quiescent
times}. In another paper (in preparation) we discuss the
correlation between pulses and neighboring intervals within the internal shock
model.

\section*{Acknowledgment}

This research was supported by a US-Israel BSF grant.


\begin{thebibliography}{10}
\bibitem{3}Fenimore E. E., Ramirez-Ruiz E., 2000 submitted to ApJ (astro-ph/0004176)
\bibitem{2}Fenimore E. E., Ramirez E.,Sumner M. C., 1997 in: Gamma-Ray Bursts, 4th Huntsville
Symposium , Meegan, C., Preece, R., \& Koshut, T., Eds., AIP Conf. Proc. NY
p. 657
\bibitem{4}Kobayashi S., Piran T., Sari R., 1997, Apj, 490, 92
\bibitem{7}Koshut T. et al., ApJ, 1995, 452, 145
\bibitem{5}Kouveliotou C., Meegan C. A., Fishman G. J., Bhat N. P., Briggs M. S., Koshut
T. M., Paciesas W. S., Pendleton G. N., 1993, ApJ, 413, L101
\bibitem{5}Li H., Fenimore E., 1996, ApJ, 469, L115
\bibitem{6}McBreen B., Hurley K. J., Long R., Metcalfe L., 1994, MNRAS, 271, 662
\bibitem{8}Nakar E., Piran T., 2001a, to appear in A\&As, proceedings of the second Rome
GRB meeting, astro-ph/0103011
\bibitem{9}Nakar E., Piran T., 2001b submitted to MNRAS, astro-ph/0103192
\bibitem{6}Norris J. P., Nemiroff R. J., Bonnell J. T., Scargle J. D., Kouveliotou C.,
Paciesas W. S., Meegan C. A., Fishman G. J., 1996, ApJ, 459, 393
\bibitem{9}Norris J. P., Marani G. F., Bonnell J. T., 2000, ApJ, 534, 248
\bibitem{11}Quilligan F., Hurley K. J., McBreen B., Hanlon L., Duggan P., 1999, A\&AS, 138,
419
\bibitem{14}Ramirez-Ruiz E., Fenimore E. E., 2000, ApJ, 539, 712
\bibitem{15}Ramirez-Ruiz E., Merloni A., 2001, MNRAS, 320, L25
\bibitem{15}Ramirez-Ruiz E., Merloni A., Rees M. J., 2001, MNRAS, 324, 1147
\bibitem{16}Reichart D. E., Lamb D. Q., Fenimore E. E., Ramirez-Ruiz E., Cline T. L., Hurley
K., 2001, ApJ, 552, 57
\bibitem{18}Sari R., Piran T., 1997, ApJ, 485, 270
\bibitem{15}Stern B., Poutanen J., Svensson R., 1997, ApJ, 489, L41
\bibitem{16}Stern B., Poutanen J., Svensson R., 1999, ApJ, 510, 312\end{thebibliography}
\end{document}